\documentclass{Interspeech}

\interspeechcameraready

\title{Contextualized Automatic Speech Recognition with Dynamic Vocabulary Prediction and Activation}

\author[affiliation={1}{,*}]{Zhennan}{Lin}
\author[affiliation={1}{,*}]{Kaixun}{Huang}
\author[affiliation={2}{,*}]{Wei}{Ren}
\author[affiliation={2}]{Linju}{Yang}
\author[affiliation={1}{,\dagger}]{Lei}{Xie}

\affiliation{Audio, Speech and Language Processing Group (ASLP@NPU), School of Computer Science}{Northwestern Polytechnical University}{Xi’an, China}
\affiliation{}{Chongqing Changan Automobile Co., Ltd.}{China}
\email{znlin@mail.nwpu.edu.cn, lxie@nwpu.edu.cn}

\UseRawInputEncoding
\usepackage{comment}

\usepackage{multirow}
\usepackage{subfig}
\usepackage{float}

\begin{document}

\maketitle

\renewcommand{\thefootnote}{\fnsymbol{footnote}}
\footnotetext[1]{Co-first-author.}
\footnotetext[2]{Corresponding author.}
\renewcommand{\thefootnote}{\arabic{footnote}}

\begin{abstract}

    Deep biasing improves automatic speech recognition (ASR) performance by incorporating contextual phrases. However, most existing methods enhance subwords in a contextual phrase as independent units, potentially compromising contextual phrase integrity, leading to accuracy reduction. In this paper, we propose an encoder-based phrase-level contextualized ASR method that leverages dynamic vocabulary prediction and activation. We introduce architectural optimizations and integrate a bias loss to extend phrase-level predictions based on frame-level outputs. We also introduce a confidence-activated decoding method that ensures the complete output of contextual phrases while suppressing incorrect bias. Experiments on Librispeech and Wenetspeech datasets demonstrate that our approach achieves relative WER reductions of 28.31\% and 23.49\% compared to baseline, with the WER on contextual phrases decreasing relatively by 72.04\% and 75.69\%.
    
\end{abstract}

\section{Introduction}

    In recent years, driven by advances in neural networks, end-to-end automatic speech recognition (E2E-ASR) has made remarkable progress~\cite{graves2006connectionist,graves2012sequence,chorowski2015attention,gulati20_interspeech,chan2016listen,Cho2014LearningPR}. However, E2E-ASR models depend heavily on their training data, resulting in a significant drop in recognition accuracy when encountering rare phrases (e.g., entity names and technical terms) in unseen contexts. Therefore, improving E2E-ASR through deep biasing is crucial for correctly recognizing rare phrases.

    To address this challenge, a typical bias method is the shallow fusion~\cite{williams2018contextual,chen2019end,zhao2019shallow,huang20f_interspeech,kim2021improved}, which uses a weighted finite-state transducer (WFST) to construct a contextual decoding graph to improve the recognition of contextual phrases. However, the improvement in contextual phrase prediction achieved through this method is limited.
    The neural network-based deep biasing method~\cite{chen2019joint,han2021cif,sun2021tree,dingliwal2023personalization,9747101} provides a better solution, by integrating a dedicated biasing module into the end-to-end model, enabling rapid adaptation to diverse scenarios through an editable list of contextual phrases. Compared to shallow fusion, deep biasing offers greater adaptability and significantly improves rare phrase recognition.

    To improve the effectiveness of contextualized ASR, previous studies introduce the biasing module for deep bias, such as CLAS~\cite{pundak2018deep} and CATT~\cite{chang2021context}. Some approaches, such as CPPN~\cite{huang23d_interspeech}, introduce additional bias loss functions to guide the model in capturing contextual information, thus improving the recognition accuracy of contextual phrases.
    However, most existing bias methods represent contextual phrases as subword sequences and with isolated optimizations to each subword. For instance, a personal name like ``Alexander'' may be segmented into the subword sequence ``A'', ``lex'', and ``ander'', where the contextualized model learns to increase the probability of the entire sequence. However, these methods often ignore the integrity of subword sequences, resulting in partial or incorrect subword predictions.
    Previous studies have attempted to address this issue using prefix tree-based methods~\cite{sun2021tree,dingliwal2023personalization,le21_interspeech} or by integrating additional text data~\cite{le21_interspeech,qiu2023improving}. These methods demand extra computational resources and face practical limitations regarding data requirements. A recent study proposes considering contextual phrases as discrete labels and introducing phrase-level bias tokens into inference using a dynamic vocabulary~\cite{Sudo2024ContextualizedAS}. Each token represents an entire contextual phrase, enabling the model to capture intricate dependencies among subwords within the phrase. However, this method is primarily implemented on the decoder, making it dependent on autoregressive decoding and less scalable than encoder-based approaches.

    Inspired by Sudo's work~\cite{Sudo2024ContextualizedAS}, we propose a contextual ASR method based on an encoder architecture that leverages dynamic vocabulary prediction and activation. We use a dynamic vocabulary and treat contextual phrases as unified tokens, integrating them into the frame-level output of the CTC model. This design enhances subword dependency modeling while reducing unintended bias. We optimize the network architecture and incorporate a bias loss function, which enhances the model's ability to learn and model contextual phrases. Furthermore, to better leverage bias tokens for improving model prediction outcomes, we propose a confidence-activated decoding strategy that incorporates CTC posterior probabilities to post-process the model outputs, thereby enabling the accurate replacement of bias tokens and their corresponding text with contextual phrases.
    Our proposed approach achieves relative reductions in WER of 28.26\% and 23.49\% on LibriSpeech~\cite{7178964} and WenetSpeech~\cite{zhang2022wenetspeech}, respectively, while the WER for contextual phrases is reduced by 71.16\% and 75.69\%, respectively. These results confirm the effectiveness of our approach, demonstrating performance improvements in popular English and Chinese speech recognition benchmarks.

\section{Method}

    This section introduces the proposed method based on the CTC architecture. To preserve the original recognition accuracy and facilitate the convergence of the bias module, we integrate the bias module into the pre-trained ASR model while keeping the original model parameters fixed, training only those associated with the bias module.

    \begin{figure}[t]
        \centering
        \begin{minipage}{\linewidth}
            \centering
            \includegraphics[width=0.9\linewidth]{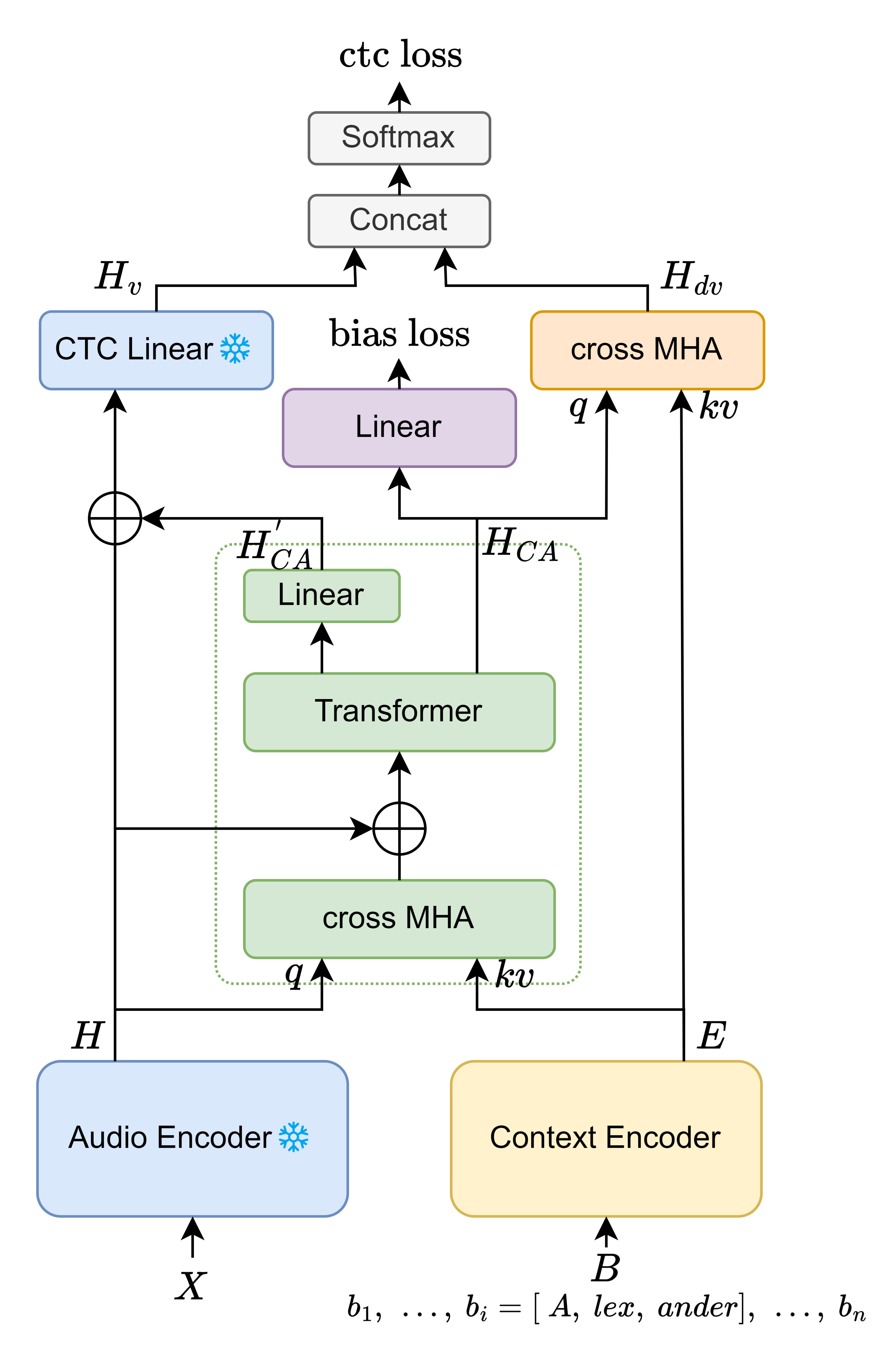}
        \end{minipage}
        
        \caption{The overall architecture of the proposed method.}
        \label{fig:framework}
        \vspace{-0.5cm} 
    \end{figure}
    
\subsection{Model structure}

    As shown in Fig.\ref{fig:framework}, to ensure the scalability of the model, we exclusively incorporate the bias module into the encoder. Building upon the CTC model architecture, we introduced a context encoder, a bias-aware module, a multi-head attention-based output layer, and a bias projection layer.

    The audio encoder transforms the input audio features $X$ into a sequence of the hidden state features $H=[h_{1},\dots,h_{T}]\in\mathbb{R}^{d\times T}$, while the context encoder converts the bias list $B$ into bias embeddings $E=[e_{1},\dots,e_{n}]\in\mathbb{R}^{d\times n}$, where $n$ represents the number of contextual phrases in the bias list. We employ Conformer as the context encoder and extract the vector at the first position of the label in the embedding sequence of each contextual phrase $b_{i}$ as the embedding vector $e_{i}$.

    Given the hidden state feature features $H$ and the bias embeddings $E$, the bias-aware module captures the relationship between the audio and the bias list to generate a contextualized high-level hidden representation $H_{CA}$. This module consists of a multi-head attention layer, a transformer, and a linear layer. The multi-head attention layer uses $H$ as the query while $E$ serves as the key and the value, producing a hidden representation with contextual information. The transformer then processed this representation to extract a high-level representation $H_{CA}$. The process can be formulated as:
    \begin{align}
    \label{eq:bias-aware module}
        H^{'} &= \text{MHA} \left( H,E,E \right), \\
        H_{CA} &= \text{Transformer} \left( H+H^{'} \right), \\
        H^{'}_{CA} &= \text{Linear} \left( H_{CA} \right).
    \end{align}
    
    Subsequently, we use the hidden representation $H_{CA}$ as the input to the output layer, while the bias embeddings $E$ serve as the key and value to compute the attention scores $H_{dv}\in\mathbb{R}^{n\times T}$, indicating the extent to which each frame attends to the contextual phrases in $B$. We average the multi-head attention scores to leverage different subspaces of attention and enhance contextual representation effectively. In addition, $H^{'}_{CA}$ is integrated into the hidden state feature sequence $H$ as the input to the CTC projection layer. This layer projects the sequence onto a vector sequence $H_{v}\in\mathbb{R}^{V\times T}$, where $V$ denotes the number of tokens in the vocabulary. The formulation is as follows:
    \begin{align}
    \label{eq:output layer}
        H_{v} &= \text{Linear} \left( H+H^{'}_{CA} \right), \\
        H_{dv} &= \frac{\text{Linear} \left( H_{CA} \right) \text{Linear} \left( E^{T} \right)}{\sqrt{d_{k}}}.
    \end{align}

    To expand the dynamic vocabulary, $H_{dv}$ is concatenated with $H_{v}$ and then passed through a softmax layer, yielding a posterior probability distribution of dimension $V+n$, which is used as the CTC decoding graph. This enables the model to incorporate $n$ additional predicted labels beyond the $V$ vocabulary tokens, representing the $n$ contextual phrases in $B$.

    To enhance the model’s ability to capture the bias list and improve its attention to contextual information, we introduce a bias loss function for joint training. Taking the original transcription $y=[\dots,A,lex,ander,\dots]$, if the sequence contains only the contextual phrase $b_{i}$, the corresponding bias loss label is set as $[A,lex,ander]$. The joint loss function is defined as follows:
    \begin{align}
    \label{eq:joint loss}
        \mathcal{L}_{total} &= \lambda_{1}\mathcal{L}_{ctc} + \lambda_{2}\mathcal{L}_{bias}.
    \end{align}
    where $\lambda_{1}$ and $\lambda_{2}$ are weight hyperparameters we set as 0.3 and 0.05 in experiment.
    
\subsection{context label strategies}

    To replace subword sequences with frame-level bias labels and ensure complete recognition of contextual phrases, we explore two strategies for contextual labeling.
    
    Consider the bias list $B$ in Fig.\ref{fig:framework}, assume the original transcription $y=[\dots,A,lex,ander,\dots]$, containing only the i-th contextual phrase $b_{i}=[A,lex,ander]$. The first word-by-word replacement strategy (WR) replaces each subword in the contextual phrase with the bias label $<b_{i}>$. Thus, the target sequence for CTC prediction is modified to $[\dots,<b_{i}>,<b_{i}>,<b_{i}>,\dots]$. The bias token indicates that the label appearing here belongs to this contextual phrase. During the inference, the CTC label merging rule is applied to the target sequence predicted by the model, and consecutive bias tokens are merged, modifying the label sequence to $[\dots,<b_{i}>,\dots]$.
    
    We discuss the second tail addition strategy (TA), in which a bias token is inserted after the target contextual phrase. Specifically, the target sequence for CTC prediction is modified to $[\dots,A,lex,ander,<b_{i}>,\dots]$, where the bias label $<b_{i}>$ serves as an indicator that the preceding tokens may correspond to a contextual phrase. Once the model predicts a bias token, we assess the prediction's reliability using a confidence-based method that incorporates the model’s acoustic information. If the confidence score exceeds a predefined threshold, the last $j$ labels can be replaced with the corresponding contextual phrase. For instance, if the predicted sequence $y'=[\dots,A,lx,ander,<b_{i}>,\dots]$, decoding triggered by the confidence threshold can refine it to $[\dots,A,lex,ander,\dots]$.

\subsection{confidence activation}

    \begin{figure}[t]
        \centering
        \begin{minipage}{\linewidth}
            \centering
            \subfloat[CTC peaks corresponding to labels]{
                \includegraphics[width=\linewidth]{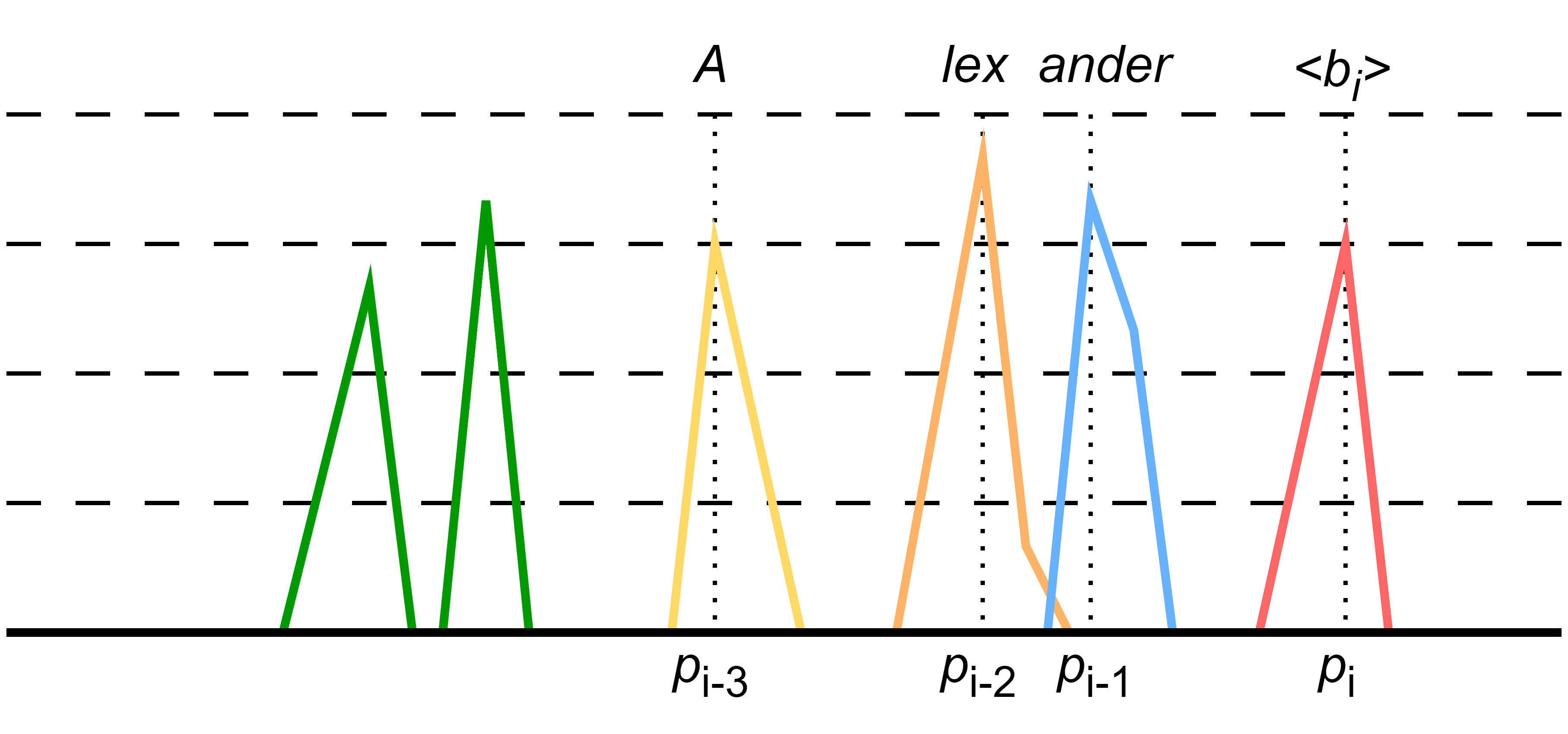}
            }
        \end{minipage}
        \vspace{0.3cm}
        \begin{minipage}{\linewidth}
            \centering
            \subfloat[Confidence activation decoding graph]{
                \includegraphics[width=\linewidth]{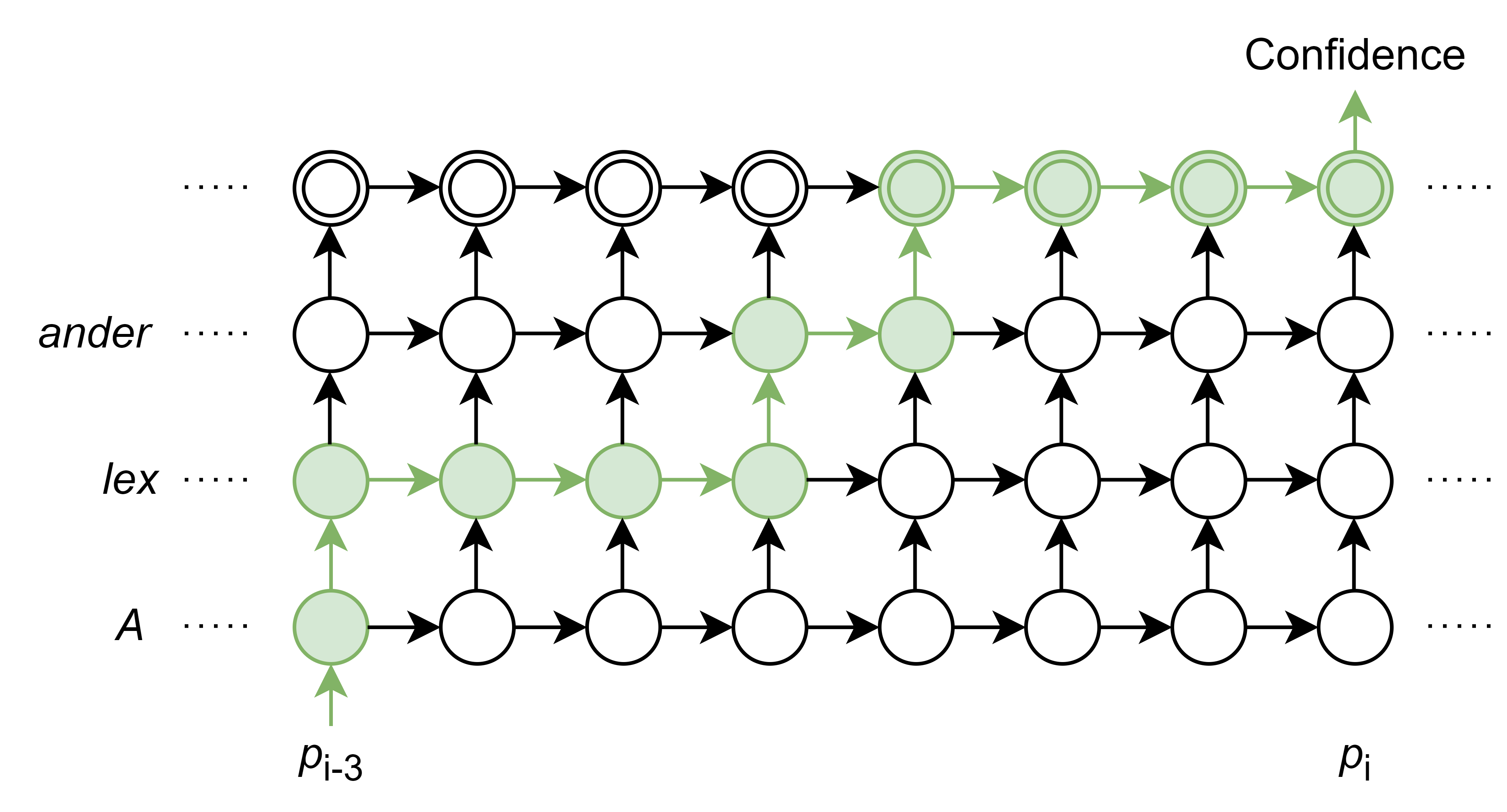}
            }
        \end{minipage}
        
        \caption{The confidence-activated decoding method.}
        \label{fig:search}
        \vspace{-0.5cm} 
    \end{figure}
    
    Our proposed confidence-activated decoding method is illustrated in Fig.\ref{fig:search}. As shown in Fig.\ref{fig:search}(a), when the model predicts the sequence $y'=[\dots, A,lex,ander,<b_{i}>,\dots]$, we identify the termination frame $p_{i}$ as the CTC peak corresponding to the label $<b_{i}>$ and determine the starting frame $p_{i-j}$ as the j-th CTC peak before $<b_{i}>$. Within the posterior probabilities spanning from the starting to the termination frame, we search for a path corresponding to the contextual phrase sequence and use the highest probability along this path as the confidence score, as shown in Fig.\ref{fig:search}(b). The confidence threshold is set as $k_{i}*threshold$, where $k_{i}$ denotes the length of the contextual phrase $b_{i}$, ensuring robust performance across varying contextual phrase lengths. Furthermore, to mitigate substitution range variations due to insertion or deletion errors, we define $j$ within the range $[k_{i}-2,k_{i}+2]$ to find the sequence with the highest confidence score. If the maximum confidence surpasses the confidence threshold, the identified sequence is replaced with the predicted phrase.

\section{Experiments}

\subsection{Experimental setup}
    We train CTC/AED models as the baseline and the pre-trained ASR model, using the Wenet toolkit~\cite{Yao2021WeNetPO}. The model takes an 80-dimensional Mel filterbank as input, with a frameshift of \SI{10}{\milli\second} and a frame length of \SI{25}{\milli\second}. SpecAugment~\cite{park19e_interspeech} is applied to enhance robustness. The audio encoder is a 12-layer Conformer, featuring an input dimension of 256, 4 attention heads, 2048 linear units, and 15 convolution kernel size. The decoder is a bi-transformer composed of 4 attention heads, 2048 linear units, 3 forward blocks, and 3 reverse blocks.
    We freeze the pre-trained model parameters and optimize only the bias module. The context encoder is a 6-layer Conformer composed of 1024 linear units. The bias-aware module comprises a 4-head attention layer and a 2-layer transformer. The extended output layer also includes a 4-head attention layer.
    During training, the batch size is set at 20, and a bias list is randomly generated for each batch. With an 80\% probability, 2 to 10 words are randomly chosen from each data sample to serve as contextual phrases.

    To demonstrate the adaptability of the proposed approach, we conduct experiments on the Librispeech~\cite{7178964} and Wenetspeech~\cite{zhang2022wenetspeech} corpora. The Librispeech dataset consists of approximately 1,000 hours of English read speech. We train the baseline and the bias module on the 960-hour train set, using the dev-clean and dev-other subsets for validation, and the test-clean and test-other subsets for evaluation. We use the bias list from the Librispeech provided in~\cite{le21_interspeech}, which constructs bias lists of sizes 100, 500, and 1,000. The distractor contextual phrases are randomly sampled from the rare vocabulary of the entire corpus.
    The Wenetspeech contextual biasing dataset contains approximately 1,000 hours of data from the Wenetspeech subset~\cite{xu2023cb}. We train the baseline and the bias module on this dataset and evaluate the results using its test set. The test set is categorized into three subsets: person, place, and organization. Named entities are extracted from the labels using the open-source toolkit HanLP1, retaining only those that appear between 5 and 700 times. A total of 298 named entities are selected to construct the bias lists for evaluation.

    In addition to evaluating ASR performance using word error rate (WER), we also employ biased word error rate (B-WER) and unbiased word error rate (U-WER) to assess the model's effectiveness in recognizing words from the contextual bias list. U-WER quantifies errors on words not included in the bias list, whereas B-WER focuses on errors involving words within the bias list. Insertion errors are categorized based on their presence in the bias list. if an inserted phrase appears in the bias list, it is counted towards B-WER, otherwise, it is counted towards U-WER.
    For the Mandarin speech dataset, we evaluate the ASR performance using CER, U-CER, and B-CER metrics.

\subsection{Analysis of context label strategies}

    \begin{table}[t]
        \caption{WER(\%) Results of different context label strategies obtained on Wenetspeech. Reported metrics are in the following format: WER(U-WER/B-WER)}
        \label{tab:prediction labels}
        \centering
        \begin{tabular}{ccc}
        \toprule
        \textbf{Strategy} & \textbf{Librispeech} & \textbf{Wenetspeech} \\
        \midrule
        baseline & \begin{tabular}[c]{@{}c@{}}9.5\\(6.68/34.3)\end{tabular} & \begin{tabular}[c]{@{}c@{}}13.28\\(10.67/26.7)\end{tabular} \\
        \midrule
        WR & \begin{tabular}[c]{@{}c@{}}\textbf{6.44}\\(\textbf{6.09}/\textbf{9.39})\end{tabular} & \begin{tabular}[c]{@{}c@{}}12.57\\(10.41/23.66)\end{tabular} \\
        \midrule
        TA & \begin{tabular}[c]{@{}c@{}}6.81\\(6.48/9.59)\end{tabular} & \begin{tabular}[c]{@{}c@{}}\textbf{10.16}\\(\textbf{10.87}/\textbf{6.49})\end{tabular} \\
        \bottomrule
        \end{tabular}
    \end{table}

    \begin{table*}[th]
        \caption{WER(\%) Results of different systems obtained on Librispeech. Reported metrics are in the following format: WER(U-WER/B-WER)}
        \label{tab:WER Results on Librispeech}
        \centering
        \begin{tabular}{ccccccccc}
        \toprule
        \multirow{2}{*}{\textbf{Model}} & \multicolumn{2}{c}{\textbf{\quad w/o bias module}} & \multicolumn{2}{c}{\textbf{\quad N=0}} & \multicolumn{2}{c}{\textbf{\quad N=100}} & \multicolumn{2}{c}{\textbf{\quad N=1000}} \\
         & \textbf{WER} & \textbf{rB-WER} & \textbf{WER} & \textbf{rB-WER} & \textbf{WER} & \textbf{rB-WER} & \textbf{WER} & \textbf{rB-WER} \\
        \midrule
        CPPN~\cite{huang23d_interspeech} & \begin{tabular}[c]{@{}c@{}}8.88\\(5.6/37.6)\end{tabular} & - & \begin{tabular}[c]{@{}c@{}}9.16\\(5.9/37.5)\end{tabular} & -0.27\% & \begin{tabular}[c]{@{}c@{}}7.77\\(6.0/23.0)\end{tabular} & -38.82\% & \begin{tabular}[c]{@{}c@{}}8.75\\(6.9/25.3)\end{tabular} & -32.71\% \\
        \midrule
        DV~\cite{Sudo2024ContextualizedAS} & \begin{tabular}[c]{@{}c@{}}5.98\\(4.0/23.1)\end{tabular} & - & \begin{tabular}[c]{@{}c@{}}6.95\\(4.6/27.5)\end{tabular} & 19.05\% & \begin{tabular}[c]{@{}c@{}}4.63\\(4.3/7.1)\end{tabular} & -69.26\% & \begin{tabular}[c]{@{}c@{}}4.97\\(4.6/8.5)\end{tabular} & -63.20\% \\
        \midrule
        DVPA-CTC & \begin{tabular}[c]{@{}c@{}}9.5\\(6.68/34.3)\end{tabular} & - & \begin{tabular}[c]{@{}c@{}}9.53\\(6.66/34.65)\end{tabular} & 1.02\% & \begin{tabular}[c]{@{}c@{}}6.81\\(6.48/9.59)\end{tabular} & -72.04\% & \begin{tabular}[c]{@{}c@{}}8.06\\(7.18/15.74)\end{tabular} & -54.11\% \\
        \midrule
        DVPA-AED & \begin{tabular}[c]{@{}c@{}}8.77\\(6.19/31.38)\end{tabular} & - & \begin{tabular}[c]{@{}c@{}}8.72\\(6.14/31.4)\end{tabular} & 0.06\% & \begin{tabular}[c]{@{}c@{}}6.12\\(5.75/9.33)\end{tabular} & -70.27\% & \begin{tabular}[c]{@{}c@{}}7.39\\(6.5/15.2)\end{tabular} & -51.56\% \\
        \bottomrule
        \end{tabular}
    \end{table*}
    

    To assess the impact of two different context label strategies on improving the recognition of contextual phrases, we compared their performance on the LibriSpeech test-other dataset (with a bias list size of 100) and the WeNetSpeech test set, evaluating their effectiveness in both English and Chinese.

    As shown in Table~\ref{tab:prediction labels}, both strategies effectively enhance the recognition of contextual phrases. The WR strategy slightly outperforms the TA strategy on the test-other dataset, but its performance is noticeably worse on the WeNetSpeech test set, particularly when dealing with longer Chinese phrases. This may be due to its limited ability to model long contextual phrases. By appending bias tags at the end, the TA strategy primarily emphasizes the similarity between contextual phrases and overall pronunciation. Given the overall results on both English and Chinese test sets, we primarily adopt the TA strategy in our subsequent experiments.


    

\subsection{Results on Librispeech}


    This section analyzes the performance of different ASR systems on the Librispeech test-other dataset under varying bias list sizes. Given the differences in the performance of the baseline, we focus on comparing the relative improvements in B-WER across systems. As shown in Table~\ref{tab:WER Results on Librispeech}, our proposed method reduces WER by 28.32\% and B-WER by 72.04\% relative to the baseline. When $N=100$, our method outperforms other contextualized ASR models, achieving notable performance gains. Although model performance declines as the bias list size increases, the overall effectiveness remains strong. Notably, even when the bias list size is set to zero, the bias module has no noticeable adverse effect on model performance.
    
\subsection{Analysis of ablation}

    \begin{table}[t]
        \caption{Ablation analysis}
        \label{tab:ablation}
        \centering
        \begin{tabular}{lccc}
        \toprule
        \multicolumn{1}{c}{\multirow{2}{*}{\textbf{Model}}} & \multicolumn{3}{c}{\textbf{Librispeech N=100}} \\
         & \textbf{WER} & \textbf{U-WER} & \textbf{B-WER} \\
        \midrule
        DVPA-CTC (Proposed) & 6.81 & 6.48 & 9.59 \\
        ~- context conformer Enc & 6.89 & 6.46 & 10.65 \\
        ~- bias loss & 7.27 & 6.77 & 11.61 \\
        ~- bias-aware module & 7.32 & 6.55 & 14.15 \\
        ~- confidence activation & 10.31 & 8.88 & 22.88 \\
        \bottomrule
        \end{tabular}
    \end{table}

    To assess the impact of individual modules on overall model performance, we conducted ablation experiments on the Librispeech dataset. The results indicate that removing the bias loss function degrades model performance. Compared with the WA strategy, the TA strategy yields a more noticeable improvement by incorporating the bias loss function for supervision. This may be because the TA strategy primarily emphasizes overall pronunciation but does not model subword pronunciation details adequately, thereby limiting its effectiveness. Additionally, removing the bias-aware module results in higher WER and B-WER, demonstrating that the component is important for effectively capturing bias information. When the confidence-activated decoding strategy is disabled, incorrect predictions and replacements have a negative impact on recognition performance, further validating the effectiveness of the confidence-activated strategy.
    
    
\subsection{Results on Wenetspeech}

    \begin{table}[t]
        \caption{WER(\%) Results obtained on Wenetspeech. Reported metrics are in the following format: WER(U-WER/B-WER)}
        \label{tab:WER Results on Wenetspeech}
        \centering
        \setlength{\tabcolsep}{4pt}
        \begin{tabular}{cccc}
        \toprule
        \textbf{Model} & \textbf{Organization} & \textbf{Person} & \textbf{Place} \\
        \midrule
        \begin{tabular}[c]{@{}c@{}}w/o\\ bias module\end{tabular} & \begin{tabular}[c]{@{}c@{}}9.36\\ (8.72/11.6)\end{tabular} & \begin{tabular}[c]{@{}c@{}}14.79\\ (10.71/33.21)\end{tabular} & \begin{tabular}[c]{@{}c@{}}13.02\\ (11.06/26.24)\end{tabular} \\
        \midrule
        \begin{tabular}[c]{@{}c@{}}w/\\ bias module\end{tabular} & \begin{tabular}[c]{@{}c@{}}7.53\\ (8.83/2.94)\end{tabular} & \begin{tabular}[c]{@{}c@{}}9.97\\ (10.65/6.9)\end{tabular} & \begin{tabular}[c]{@{}c@{}}10.31\\ (10.98/5.82)\end{tabular} \\
        \bottomrule
        \end{tabular}
    \end{table}
    
    To evaluate the effectiveness of the approach on Chinese, we conduct experiments on the WeNetSpeech test set. As shown in Table~\ref{tab:WER Results on Wenetspeech}, the results indicate that our proposed method improves WER and B-WER performance in Chinese tasks while demonstrating robustness in handling contextual phrases of various types and lengths (e.g., organization names, place names, and personal names). Specifically, our model achieved an average 23.49\% relative improvement in CER and a 75.69\% relative improvement in B-CER. These results further illustrate the effectiveness of the proposed model for contextualized ASR tasks in Chinese.
    
\section{Conclusion}

    In this paper, we propose a contextual deep biasing approach for speech recognition that leverages dynamic vocabulary prediction and activation. We investigate two labeling strategies to adaptively propagate phrase-level labels to frame-level outputs. To ensure the integrity of subword sequences while mitigating excessive boosting of contextual phrases, we introduce a confidence-activated decoding method. Additionally, we refine the network architecture by incorporating a bias loss, encouraging the model to capture dependencies among subwords more effectively. Experimental results demonstrate that our method outperforms previous approaches in both Chinese and English speech recognition.

\bibliographystyle{IEEEtran}
\bibliography{mybib}

\end{document}